\documentclass[12pt]{iopart}

\usepackage{amsfonts, amssymb, amsgen, amsthm}
\usepackage{multicol}
\usepackage{braket}

\usepackage{graphicx}

\usepackage{comment}

\usepackage{cite}

\usepackage{bm}

\newcommand{\eq}[1]{\begin{equation} #1 \end{equation}}
\newcommand{\eqa}[2]{\begin{equation} #1 \label{#2} \end{equation}}
\newcommand{\balign}[1]{\begin{eqnarray} #1 \end{eqnarray}}

\newcommand{\figin}[4]
{\begin{figure}[tb]
\centering
\includegraphics[width= #1]{#2.pdf}
\caption{#3}
\label{f:#4}
\end{figure}}

\newcommand{\todayd}{\the\year/\the\month/\the\day}
\newcommand{\del}{\partial}
\newcommand{\bib}{\bibitem}

\newcommand{\lmd}{\lambda}

\newcommand{\lb}{\label}
\newcommand{\nt}{\nonumber}
\renewcommand{\fref}[1]{Fig.~\ref{f:#1}}
\newcommand{\eqref}[1]{(\ref{#1})}
\renewcommand{\sref}[1]{Sec.~\ref{s:#1}}

\newcommand{\bel}{\begin{easylist}}
\newcommand{\eel}{\end{easylist}}

\def \({\left(}
\def \){\right)}
\newcommand{\la}{\left\langle}
\newcommand{\ra}{\right\rangle}
\def \[{\left[}
\def \]{\right]}
\newcommand{\abs}[1]{\left|#1\right|}

\newcommand{\sumtwo}[2]%
{\mathop{\sum_{#1}}_{#2}}
\newcommand{\sumthree}[3]%
{\mathop{\mathop{\sum_{#1}}_{#2}}_{#3}}
\newcommand{\sumfour}[4]%
{\mathop{\mathop{\mathop{\sum_{#1}}_{#2}}_{#3}}_{#4}} 
\newcommand{\prodtwo}[2]%
{\mathop{\prod_{#1}}_{#2}}
\newcommand{\mintwo}[2]%
{\mathop{\min_{#1}}_{#2}}
\newcommand{\maxtwo}[2]%
{\mathop{\max_{#1}}_{#2}}
\newcommand{\maxthree}[3]%
{\mathop{\mathop{\max_{#1}}_{#2}}_{#3}}
\newcommand{\limtwo}[2]%
{\mathop{\lim_{#1}}_{#2}}
\newcommand{\suptwo}[2]%
{\mathop{\sup_{#1}}_{#2}}
\newcommand{\supthree}[3]%
{\mathop{\mathop{\sup_{#1}}_{#2}}_{#3}}
\newcommand{\supfour}[4]%
{\mathop{\mathop{\mathop{\sup_{#1}}_{#2}}_{#3}}_{#4}} 
\newcommand{\inftwo}[2]%
{\mathop{\inf_{#1}}_{#2}}
\newcommand{\infthree}[3]%
{\mathop{\mathop{\inf_{#1}}_{#2}}_{#3}}
\newcommand{\inffour}[4]%
{\mathop{\mathop{\mathop{\inf_{#1}}_{#2}}_{#3}}_{#4}} 

\newcommand\calD{{\cal D}}

\newcommand\calJ{{\cal J}}

\newcommand\calP{{\cal P}}

\newcommand\calW{{\cal W}}

\newcommand{\bsj}{\boldsymbol{j}}

\newcommand{\bsp}{\boldsymbol{p}}

\newcommand{\bsu}{\boldsymbol{u}}
\newcommand{\bsv}{\boldsymbol{v}}

\newcommand{\bbR}{\mathbb{R}}

\newcommand{\ep}{\varepsilon}

\newcommand{\Di}{\mathit{\Delta}}

\newcommand{\tlr}{\tilde{R}}

\newcommand{\dsgm}{\dot{\sigma}}

\newcommand{\pss}{p^{\rm ss}}

\newcommand{\dPi}{\dot{\Pi}}

\newcommand{\dsgmex}{\dsgm^{\rm MN}}

\newcommand{\dsgmhk}{\dsgm^{\rm MN, hk}}

\def\rnum#1{\resizebox{0.5em}{\height}{\expandafter{\romannumeral #1}}}
\def\Rnum#1{\resizebox{0.5em}{\height}{\uppercase\expandafter{\romannumeral #1}}}

\newtheorem{thm}{Theorem}
\newtheorem{lm}[thm]{Lemma}
\newtheorem{pro}[thm]{Proposition}

\newcommand{\bpf}[1]{\begin{proof} #1 \end{proof}}

\newcommand{\bthm}[1]{\begin{thm} #1 \end{thm}}

\newcommand{\bpro}[1]{\begin{pro} #1 \end{pro}}

\theoremstyle{definition}
\newtheorem{dfn}{Definition}
\newcommand{\bdf}[1]{\begin{dfn} #1 \end{dfn}}

\begin{document}

\title{Wasserstein distance in speed limit inequalities for Markov jump processes}

\author{Naoto Shiraishi}

\address{ Faculty of arts and sciences, University of Tokyo, 3-8-1 Komaba, Meguro-ku, Tokyo, Japan}
\ead{shiraishi@phys.c.u-tokyo.ac.jp}
\vspace{10pt}

\begin{abstract}
The role of the Wasserstein distance in the thermodynamic speed limit inequalities for Markov jump processes is investigated.
We elucidate the nature of the Wasserstein distance in the thermodynamic speed limit inequality from three different perspectives with resolving three remaining problems.
In the first part, we derive a unified speed limit inequality for a general weighted graph, which reproduces both the conventional speed limit inequality and the trade-off relation between current and entropy production as its special case.
In the second part, we treat the setting where the tightest bound with the Wasserstein distance has not yet been obtained and investigate why such a bound is out of reach.
In the third part, we compare the speed limit inequalities for Markov jump processes with L$^1$-Wasserstein distance and for overdamped Langevin systems with L$^2$-Wasserstein distance, and argue that these two have different origins despite their apparent similarity.
\end{abstract}

%
%
%
%
%

\tableofcontents

\section{Introduction}\label{intro}

Entropy production, defined as the sum of the entropy increase of the system and the attached baths, is a key quantity in the stochastic thermodynamics quantifying thermodynamic irreversibility of processes~\cite{Sei12, Shibook}.
The fluctuation theorem~\cite{ECM93, GC95, Jar97, Kur98, Mae99, Jar00} reveals a hidden symmetry in entropy production, which clarifies the mathematical structure of the thermodynamic irreversibility.
Notably, the fluctuation theorem and its variants reproduce various nonequilibrium relations, including the fluctuation-response relation and the Onsager reciprocity theorem, and also produce these generalizations~\cite{ES95, Gal96, PJP09, SS10, Nak11, APE16, PE17, Shi22}.
After the discovery of the fluctuation theorem and Jarzynski equality, a variety of variants of the fluctuation theorem have been proposed and investigated~\cite{HS01, SS05, SU12, KN13, SS15, PE17}.

Recent developments in stochastic thermodynamics elucidate further roles of entropy production; thermodynamic bounds.
A prominent example is the speed limit inequality~\cite{Aur12, SFS18, VVH20, VH21, Dec22, VS23, Kol22, VS23b}, where entropy production limits the maximum speed of processes, stating that a quick process must accompany large entropy production.
Another famous example is the thermodynamic uncertainty relation~\cite{BS15, Ging16, GRH17, DS20, LGU20, KS20, Shi21, DS21} and trade-off inequalities between currents and dissipation~\cite{SST16, PS18, SS19}, which manifest the incompatibility of current, current fluctuation, and entropy production.
Further examples are restrictions on possible relaxation paths~\cite{SS19b, Kol22} and bounds on fluctuation oscillations in stationary systems~\cite{OSB22, OIK23, Shi23, VVS23}.

In the early proposal of the speed limit inequalities for classical stochastic processes, the distance between the initial and the final states is measured by the total variation distance for Markov jump processes~\cite{SFS18, VVH20} and by the Wasserstein distance for overdamped Langevin systems~\cite{Aur12}.
Later, speed limit inequalities with the Wasserstein distance for Markov jump processes are proposed~\cite{VH21, Dec22, Kol22, VS23b}.
Although various metrics supply the same speed limit bound (thermodynamic length) near equilibrium~\cite{Wei75, Rup79, SB83, SNI84, SNB85, SS97, Cro07},  several studies imply that the Wasserstein distance yields a tight bound far from equilibrium~\cite{NI21, CR23}, whose thermodynamic role is also investigated~\cite{DS19, NI21}.

However, this is not the end of the investigation of the speed limit inequalities, and there still remain various unsolved problems.
We here list three problems to be resolved:
\begin{itemize}
\item The speed limit inequality is very similar to the trade-off inequality between current and dissipation~\cite{SST16, SS19}. 
However, despite its similarity the connection between them has not yet been fully clarified.
\item In the case of stationary dissipation, the Wasserstein distance is not available for the tightest form of speed limit inequalities, at least at present. 
It is unclear why the Wasserstein distance does not fit to the tightest form.
\item The Wasserstein distance has some variations, and the speed limit inequalities for Markov jump processes and overdamped Langevin systems employ different Wasserstein distances.
The relationship between these two speed limit inequalities (whether these two are essentially the same or not) has not yet been addressed.
\end{itemize}

In this paper, we tackle these three problems.
After a brief introduction in \sref{framework}, we consider these three problems in the subsequent three sections.

In \sref{SL-w}, we establish a bridge between the speed limit inequality and the trade-off relation between current and dissipation, or short-time thermodynamic uncertainty relation.
We extend the speed limit inequality with the Wasserstein distance and the pseudo entropy production proposed by Dechant~\cite{Dec22} to a weighted graph.
The obtained speed limit inequality provides a unified framework of these two types of inequalities.

In \sref{SL-excess}, we consider systems with stationary heat dissipation and seek strong inequalities by replacing entropy production by excess entropy productions, which are generalizations of entropy production.
We compare two excess entropy productions, the Hatano-Sasa excess entropy production~\cite{HS01} and the Maes-Neto\v{c}n\'{y}-type excess entropy production~\cite{Kol22}, and two distances, the Wasserstein distance and the total variation distance in terms of the tightness of speed limit inequalities.
With this regard, the combination of the Hatano-Sasa excess entropy production and the Wasserstein distance is the best pair, while a speed limit inequality with this combination has not yet been obtained.
To consider the reason, we clarify and generalize two speed limit inequalities with the second best combinations, the Hatano-Sasa excess entropy production and the total variation distance, and  the Maes-Neto\v{c}n\'{y}-type excess entropy production and the Wasserstein distance.
This sheds light on why the speed limit inequality with the strongest combination is difficult to obtain.

In \sref{Langevin}, we compare the speed limit inequality for Markov jump processes obtained in \sref{SL-w} to that for overdamped Langevin systems.
Interestingly, the speed limit inequalities for Markov jump processes employ the L$^1$-Wasserstein distance, while that for overdamped Langevin systems employs the L$^2$-Wasserstein distance.
Then, the question naturally arises whether these two speed limit inequalities are essentially the same relation or two different relations.
Dechant~\cite{Dec22} argued that these two are essentially the same despite several apparent differences.
Contrarily to this anticipation, we carefully take the continuous Langevin limit and show that the speed limit inequality for Markov jump processes under the continuous limit does not coincide with that for overdamped Langevin systems.
We also discuss the origin of this discrepancy.

\section{Basic framework}\lb{s:framework}

\subsection{Framework of stochastic thermodynamics}

The most of this paper is devoted to Markov jump processes on discrete states.
Time evolution of the probability distribution $p_i$ on state $i\in \Omega$ is given by the master equation:
\eq{
\frac{d}{dt}p_i =\sum_j R_{ij}p_j,
}
where $R_{ij}$ is the transition matrix.
The transition matrix satisfies the nonnegative condition ($R_{ij}\geq 0$ for all $i\neq j$) and the normalization condition ($\sum_i R_{ij}=0$ for all $i$).
The transition matrix $R_{ij}$ is time-dependent in general.

If a Markov jump process is induced by multiple baths, the transition matrix is decomposed into that for each bath as
\eq{
R_{ij}=\sum_\nu R_{ij}^\nu ,
}
where $\nu$ is the label of baths.
Let $\beta_\nu$ be the inverse temperature of baths, with setting the Boltzmann constant to unity.
Since a bath induces thermalization of the attached system, we require that $R^\nu$ has the Boltzmann distribution with $\beta_\nu$ as its stationary distribution: 
\eq{
\sum_j R_{ij}^\nu e^{-\beta_\nu E_j}=0,
}
defining $E_j$ as the energy of state $j$.

Many small thermal stochastic systems satisfy a stronger constraint than above, called the {\it local detailed-balance condition}:
\bdf{[Local detailed-balance condition]
A system satisfies the local detailed-balance condition if there is no microscopic current between any two states in equilibrium, which can be expressed as
\eqa{
\frac{R_{ij}^\nu}{R_{ji}^\nu}=e^{-\beta_\nu (E_i-E_j)}
}{LDB}
for all $i, j$ and all $\nu$.
}
If the system is attached to particle baths, we add the contribution of particle exchange to the exponent.

\subsection{Quantities in stochastic thermodynamics}

We shall introduce several important quantities in stochastic thermodynamics.
The central quantity in stochastic thermodynamics is the entropy production denoted by $\sigma$, quantifying the degree of thermodynamics irreversibility.

Entropy production is the sum of the entropy increase of baths and the (Shannon) entropy increase of the system.
In terms of stochastic processes, the entropy production rate (entropy production per unit time) is defined as follows.

\bdf{[Entropy production and entropy production rate]
Consider a stochastic system induced by heat baths.
The $\nu$-th bath is at inverse temperature $\beta_\nu$.
Then, the entropy production rate is defined as
\eqa{
\dsgm:=\sum_\nu \sum_{i,j} R_{ij}^\nu p_j \( \beta_\nu (E_j-E_i) +\ln \frac{p_j}{p_i}\) .
}{def-dsgm}
The entropy production is defined as its time integration:
\eq{
\sigma:=\int_0^\tau dt \dsgm.
}
}

Under the local detailed-balance condition \eqref{LDB}, the entropy production rate \eref{def-dsgm} can be expressed in a more appealing form.
In this paper, we frequently use this expression.
\bthm{
With assuming the local detailed-balance condition \eqref{LDB}, the entropy production rate is expressed as
\eqa{
\dsgm=\sum_\nu \sum_{i,j} R_{ij}^\nu p_j\ln \frac{R_{ij}^\nu p_j}{R_{ji}^\nu p_i}.
}{dsgm-ldb}
}
Under the local detailed-balance condition, we have a good lower bound of entropy production rate.
The expression \eqref{dsgm-ldb} is bounded from below as
\balign{
\dsgm=\frac12 \sum_\nu \sum_{i,j} (R_{ij}^\nu p_j-R_{ji}^\nu p_i) \ln \frac{R_{ij}^\nu p_j}{R_{ji}^\nu p_i} \geq \sum_\nu \sum_{i,j} \frac{(R_{ij}^\nu p_j-R_{ji}^\nu p_i)^2}{R_{ij}^\nu p_j+R_{ji}^\nu p_i}.
}
We call the right-hand side as {\it pseudo entropy production rate} denoted by $\dPi$~\cite{Shi21}.

\bdf{[Pseudo entropy production and pseudo entropy production rate]
Consider a system with the local detailed-balance condition.
We define the pseudo entropy production rate as
\eq{
\dPi:=\sum_\nu \sum_{i,j} \frac{(R_{ij}^\nu p_j-R_{ji}^\nu p_i)^2}{R_{ij}^\nu p_j+R_{ji}^\nu p_i}.
}
We also define {\it pseudo entropy production} as its time integration:
\eq{
\Pi:=\int_0^\tau dt \dPi.
}
}

The pseudo entropy production frequently appears in various thermodynamic bounds besides speed limit inequalities.
Examples are the thermodynamic uncertainty relation~\cite{Shi21} and the bound on fluctuation oscillation~\cite{Shi23}.
In these cases, the upper bound with entropy production comes just from the inequality $\sigma\geq \Pi$, and the pseudo entropy production serves as a true upper bound.

\bigskip

Another important quantity in stochastic thermodynamics is the activity $A$, quantifying the inherent time-scale of the system.

\bdf{[Activity]
The activity is the average number of jumps per unit time defined as
\eq{
A:=\sum_{(i,j)}R_{ij}p_j+R_{ji}p_i=\sum_\nu \sum_{(i,j)}R^\nu_{ij}p_j+R^\nu_{ji}p_i,
}
where $\sum_{(i,j)}$ is the sum over all pairs $(i,j)$ with identifying $(i,j)$ and $(j,i)$.
}

If all transition rates become $k$ times, the activity also becomes $k$ times, which meets the meaning of the time-scale of the system.
Its summand $A_{ij}:=R_{ij}p_j+R_{ji}p_i$ or $A_{ij}^\nu:=R^\nu_{ij}p_j+R^\nu_{ji}p_i$ is called {\it traffic}~\cite{MNW08, MN08}.

The activity appears in many places from glassy dynamics~\cite{Gar07, LAW07, BT12} to characterization of nonequilibrium stationary systems~\cite{MW06, Mae20b, Shi22} as a key quantity to describe processes far from equilibrium.

\subsection{Wasserstein distance}

To quantify the distance between two probability distributions, in this paper we mainly employ the Wasserstein distance, which has been studied in the context of the optimal transport theory in mathematics and recently applied to various research fields including physics~\cite{Aur12, VH21, Dec22, Kol22, VS23, VS23b} and machine learning~\cite{WGAN, WAE}.
The Wasserstein distance is a distance in the space of distributions on a metric space.
Let $\Omega$ be a set with a distance function $d$, and $\calP_\Omega$ be a space of probability distributions on $\Omega$.

\bdf{[Wasserstein distance]
The L$^c$-Wasserstein distance $\calW_c$ with $1\leq c<\infty$ between two probability distributions $p, p'\in \calP_\Omega$ is defined as
\eqa{
\calW_c(p,p'):=\min_{P\in \Pi(p,p')}\[ \sum_{i,j\in \Omega} d^c(i,j)P_{i,j}\] ^{1/c},
}{W-def}
where $d(i,j)$ is a distance between states $i$ and $j$, and $\Pi(p,p')\subseteq \calP_{\Omega\otimes \Omega}$ is a class of joint probability distributions called transport plans $P_{i,j}$, satisfying $\sum_i P_{i,j}=p_j$ and $\sum_j P_{i,j}=p'_i$.
}

If the state space $\Omega$ is a continuous space, the sum in the above definition is replaced by the integration, and probability distributions are replaced by probability density distributions.
In the space $\bbR^n$, a natural choice of the distance $d$ is the Euclid norm.

Regarding plan $P_{i,j}$ as transported mass from $j$ to $i$ and distance $d(i,j)$ as a cost for transport per unit mass, we can understand the Wasserstein distance as a minimum cost to transform $p$ to $p'$.
If a plan $P$ achieves the minimum, only one of $P_{i,j}$ or $P_{j,i}$ can take a nonzero value (Otherwise, subtraction of the smaller one always improves the cost).
In addition, in the case that $c=1$ and the change between $p$ and $p'$ is small, there exists an optimal plan $P^*$ which has nonzero value only between two neighboring vertices (i.e., $P^*_{i,j}=0$ for all $i$ and $j$ which are not directly connected by an edge).
This fact can be understood as follows: 
Consider edges $ij$ and $jk$ where $i$ and $k$ are not connected by an edge.
Suppose that a plan $P$ achieves the minimum with nonzero $P_{ik}$ and $P_{ii}$ for all $i$ are large.
Then, a modified plan $P'$ with setting $P'_{ik}=0$, $P'_{ij}=P_{ij}+P_{ik}$, $P'_{jk}=P_{jk}+P_{ik}$, and $P'_{jj}=P_{jj}-P_{ik}$ and keeping others unchanged $P'_{lm}=P_{lm}$ also achieves the minimum.
Applying this modification repeatedly, we can obtain the desired optimal plan.

A natural choice of the distance on a graph is the minimum length of the path from one vertex to another with setting the length of all edges as one.
For example, the distance between A and B in the unweighted graph on the left of \fref{Wasserstein} is two (A-B-D is the shortest path).

We can generalize this idea to general weighted graphs.
We assign a weight $w_{ij}$ on each edge $ij$ and regard it as the distance of the edge $ij$.
By setting all weights $w$ to one, we recover the original graph distance as mentioned above.
To quantify the Wasserstein distance, we minimize the length of the path in this sense.
Remark that the shortest path in the sense of an unweighted graph is not necessarily the minimizing path in a weighted graph.
For example, the distance between A and D in the weighted graph in the right of \fref{Wasserstein} is five, and the shortest path is A-C-E-D, not A-B-D.

\figin{12cm}{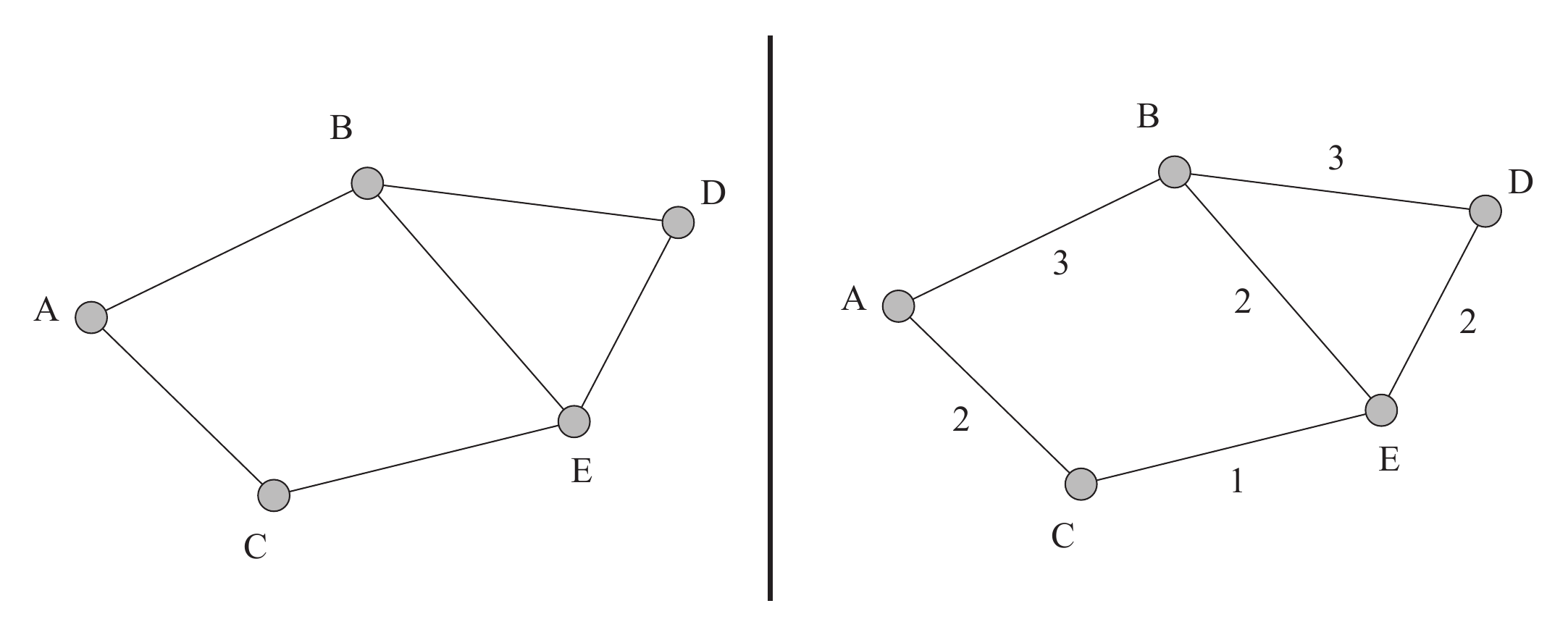}{
Left: An example of an unweighted graph.
Right: An example of a weighted graph. 
}{Wasserstein}

In the argument on Markov jump processes, we construct a graph such that vertices correspond to microscopic states, and two vertices $i$ and $j$ are connected by an edge if transition rates $R_{ij}$ or $R_{ji}$ is nonzero.


\bigskip

The significance of the L$^1$-Wasserstein distance is the Kantorovich-Rubinstein duality.
Below, we call a function $f$ is $c$-Lipschitz if $\abs{f_i-f_j}\leq c d(i,j)$ is satisfied for all $i,j$.

\bthm{[Kantorovich-Rubinstein duality]
The L$^1$-Wasserstein distance with distance function $d$ has the following expression:
\eqa{
\calW_1(p,p')=\sup_f \[ \sum_i f_i (p'_i-p_i)\] ,
}{KR-duality}
where $f$ runs all possible 1-Lipschitz functions.
}

Thanks to the Kantorovich-Rubinstein duality, we can interpret the Wasserstein distance in terms of the change in a state variable.
Let $\psi$ be the optimizer function $f$ in \eref{KR-duality}.
Then, the Wasserstein distance is equal to the change in the state variable $\psi$.
As seen later, this technique enables us to reduce trade-off inequalities on state variables to speed limit inequalities.


\section{Speed limit inequality on weighted discrete graph}\lb{s:SL-w}

\subsection{Inequality}

We start with the speed limit inequality for Markov jump processes with the Wasserstein distance shown by Dechant~\cite{Dec22}.
Here, the distance is defined in the aforementioned (unweighted) manner.

\bthm{[Dechant~\cite{Dec22}]
Consider a Markov jump process with the local detailed-balance condition whose initial and final distributions are $p(0)=p$ and $p(\tau)=p'$.
Then, the operation time $\tau$ is bounded as
\eqa{
\tau \geq \frac{2\calW_1^2(p,p')}{\overline{A}^\tau \Pi},
}{ineq-Dechant}
where $\overline{A}^\tau:=\frac1\tau \int_0^\tau dtA$ is the time average of the activity.
}

This inequality clearly shows that the pseudo entropy production is the cost of quick state transformation since the activity is the measure of time-scale.
Interestingly, this inequality achieves its equality for general $p$, $p'$, and $\tau$, which means that this inequality is tight and optimal.
In fact, various known speed limit inequalities can be directly obtained from \eref{ineq-Dechant}.
For example, the speed limit inequality shown in Ref.~\cite{SFS18}
\eq{
\tau \geq \frac{\|p-p'\|_1^2}{2\overline{A}^\tau \sigma}
}
with the total variation distance $\|v-u\|_1:=\sum_i \abs{v_i-u_i}$ can be obtained from \eref{ineq-Dechant}  by using $2\calW_1(p,p')\geq \|p-p'\|_1$ and $\Pi\leq \sigma$.
Although various definitions of distances serve as a good measure in the speed limit close to equilibrium~\cite{Wei75, Rup79, SB83, SNI84, SNB85, SS97, Cro07}, the Wasserstein distance is argued to be the best choice in the speed limit far from equilibrium~\cite{NI21, CR23}.

The original paper by Dechant~\cite{Dec22} provides just a proof outline, and a textbook~\cite{Shibook} provides its proof while this proof is complicated and not transparent.
In this paper, we generalize \eref{ineq-Dechant} with a much clearer proof:

\bthm{\lb{t:main-w}
Consider a Markov jump process with the local detailed-balance condition whose initial and final distributions are $p(0)=p$ and $p(\tau)=p'$.
Then, for any weighted graph describing this process, the operation time $\tau$ is bounded as
\eqa{
\tau \geq \frac{2\calW_{1,w}^2(p,p')}{\overline{A_w}^\tau \Pi},
}{main-w}
where $A_w:=\sum_{(i,j)}w_{ij}^2(R_{ij}p_j+R_{ji}p_i)$ is a generalization of activity to the weighted space.
}

Setting $w_{ij}=1$ for all $i,j$, we recover \eref{ineq-Dechant}.
As we shall see in the next subsection, this inequality serves as a bridge between the speed limit inequalities and the trade-off relation between currents and entropy production~\cite{SST16, SS19}, which was investigated in the context of the efficiency-power trade-off in heat engines.

Note that Vu and Saito~\cite{VS23} have recently shown a similar inequality, which employs not a modified entropy production but a modified activity but shares the essence of \eref{main-w}.
Since our aim is not only deriving the inequality but also clarifying its equality condition, we here provide pedagogical proof with detailing the constraints of optimization problems in consideration.

\bpf{
We derive the desired inequality in two steps.
First, given the average traffic $\overline{A_{w,ij}}^\tau$ and a plan $P_{ij}$ which takes a nonzero value only between neighboring states and only one of $P_{ij}$ or $P_{ji}$, we minimize the partial pseudo entropy production $\Pi_{ij}$ by changing $R_{ij}$ and $R_{ji}$.
Since this minimization problem is formulated on each single edge, we need to treat only a single edge $ij$.
Without loss of generality, we suppose that $P_{ij}>0$ and $P_{ji}=0$.

For notational simplicity, we define $j_{ij}(t):=R_{ij}(t)p_j(t)$ and $j_{ji}(t):=R_{ji}(t)p_i(t)$.
Here, we explicitly write $t$ dependence.
Our task is to minimize 
\eq{
\Pi_{ij}=\int_0^\tau dt \frac{(j_{ij}(t)-j_{ji}(t))^2}{j_{ij}(t)+j_{ji}(t)}
}
by changing $j_{ij}(t)$ and $j_{ji}(t)$ under the constraints
\balign{
\frac{\tau \overline{A_{w,ij}}^\tau}{w_{ij}^2} &=& \int_0^\tau dt (j_{ij}(t)+j_{ji}(t)), \\
P_{ij}&=&\int_0^\tau dt (j_{ij}(t)-j_{ji}(t)).
}
This optimization problem is easily solved by using the Schwarz inequality as
\balign{
\frac{\tau \overline{A_{w,ij}}^\tau}{w_{ij}^2} \Pi_{ij}&=&\( \int_0^\tau dt (j_{ij}(t)+j_{ji}(t)) \) \( \int_0^\tau dt \frac{(j_{ij}(t)-j_{ji}(t))^2}{j_{ij}(t)+j_{ji}(t)}\) \nt \\
&\geq&\( \int_0^\tau dt \abs{j_{ij}(t)-j_{ji}(t)} \) ^2 \nt \\
&\geq&\( \int_0^\tau dt (j_{ij}(t)-j_{ji}(t))\) ^2 \nt \\
&=&P_{ij}^2. \lb{wmid-key}
}
Hence, the lower bound of $\Pi_{ij}$ is
\eq{
\Pi_{ij}\geq \frac{w_{ij}^2P_{ij}^2}{\tau \overline{A_{w,ij}}^\tau}.
}
Importantly, we can achieve its equality by choosing
\balign{
j_{ij}(t)&=\frac{P_{ij}}{\tau}+\frac{\overline{A_{w,ij}}^\tau}{w_{ij}^2}, \\
j_{ji}(t)&=-\frac{P_{ij}}{\tau}+\frac{\overline{A_{w,ij}}^\tau}{w_{ij}^2}.
}

Next, we connect the above relation on each edge with the inequality on the entire system.
Recall that
\balign{
\overline{A_w}^\tau&=\sum_{(i,j)}\overline{A_{w,ij}}^\tau \\
\Pi&=\sum_{i,j}\Pi_{ij}=2\sum_{(i,j)}\Pi_{ij}
}
and
\eqa{
\calW_{1,w}(p,p')\leq \sum_{(i,j)} w_{ij}P_{ij}
}{wmid-Wasser}
for any plan $P_{ij}$ which takes a nonzero value only between neighboring states and only one of $P_{ij}$ or $P_{ji}$.
Here, the direction of $(i,j)$ is taken such that $P_{ij}$ is nonzero and $P_{ji}$ is zero.
Note that the equality in \eref{wmid-Wasser} is achievable by the definition of the Wasserstein distance.

Now we are ready to derive the desired relation
\balign{
\frac12 \tau\overline{A_w}^\tau \Pi&=&\( \sum_{(i,j)}\tau\overline{A_{w,ij}}^\tau\) \( \sum_{(i,j)} \Pi_{ij}\)  \nt \\
&\geq&\( \sum_{(i,j)}\tau\overline{A_{w,ij}}^\tau \Pi_{ij}\) ^2 \nt \\
&\geq&\( \sum_{(i,j)} w_{ij}P_{ij}\)^2 \nt \\
&\geq & \calW^2_{1,w}(p,p'),
}
where the first inequality comes from the Schwarz inequality, and the second inequality comes from \eref{wmid-key}.
The equality of the above three inequalities can be achieved simultaneously by setting 
\eq{
\overline{A_{w,ij}}^\tau=\overline{A_w}^\tau \frac{w_{ij}P^*_{ij}}{\sum_{(i,j)} w_{ij}P^*_{ij}}
}
for all $(i,j)$ with the optimal plan $P^*$.
}

\subsection{Trade-off relation between current and (pseudo-)entropy production}\lb{s:SL-tradeoff}

Consider a state variable (conserved quantity) $a$, and we investigate the trade-off relation between the current of $a$ and (pseudo) entropy production.
By setting $w_{ij}=a_i-a_j$, the Wasserstein distance is equal to the change in $a$ between the initial and the final states regardless of the plans:
\balign{
\calW_{1,w}(p,p')&=&\min_{P\in \Pi(p,p')} \[ \sum_{(i,j)} (a_i-a_j)(j_{ij}-j_{ji}) \] \nt \\
 &=&\frac12\( \sum_i a_i p'_i-\sum_i a_i p_i\) \nt \\
 &=&\frac12 \( \la a\ra_{p'} -\la a\ra_p\) ,
}
where we defined $\la a\ra_p:=\sum_i a_i p_i$.
In particular, by setting time interval $\tau$ to be small, the above quantity becomes a current $\frac12 J_a\tau$.
In this setting, the generalized activity with this $w$ reads
\eqa{
A_w=\sum_{(i,j)} (a_i-a_j)^2 (j_{ij}+j_{ji}),
}{def-Aw}
meaning the current fluctuation of $a$.
With this setting, \eref{main-w} can be read as the previously-obtained trade-off inequality between efficiency and power of heat engines obtained by Ref.~\cite{SST16}:
\eq{
\abs{J_a}\leq \sqrt{\Theta^{(2)} \dPi}\leq \sqrt{\Theta^{(2)} \dsgm},
}
where employ a symbol $\Theta^{(2)}$ following Ref.~\cite{SST16}, which is  the half of $A_w$.

In summary, \eref{main-w} serves as a unified framework capturing both the speed limit inequality and the trade-off relation between current and dissipation in a single form.



\section{Speed limit inequality with excess entropy production}\lb{s:SL-excess}

\subsection{Two excess entropy productions}

If the total system has a nonzero stationary current in its stationary state, the (pseudo) entropy production does not go to zero even in quasistatic processes.
The optimal speed limit derived in the previous section achieves its equality in the case of a single heat bath.
In this section, we investigate other types of speed limit inequalities providing a good bound even with a finite stationary current.

To this end, we consider generalizations of the entropy production such that we subtract a corresponding stationary entropy production (called {\it housekeeping entropy production}) from the total entropy production and define the remaining as {\it excess entropy production}, which truly contributes to the state conversion.
We note that the definitions of the corresponding stationary entropy production and the excess entropy production are not unique, and at least there are two definitions called the {\it Hatano-Sasa excess entropy production} and {\it Maes-Neto\v{c}n\'{y}-type excess entropy production}, which we abbreviate HS entropy production and MN entropy production, respectively.
This difference comes from what state is regarded as the corresponding stationary state:
The HS entropy production respects the transition rate of a given process and employs its stationary distribution, while the MN entropy production respects the present distribution and seeks a transition rate having the present distribution as its stationary distribution.

\bigskip

We first explain the Hatano-Sasa excess entropy production.
For a given transition rate $R$, we introduce the {\it dual transition rate} $\tlr$ as
\eq{
\tlr_{ij}:=\frac{R_{ji}\pss_i}{\pss_j},
}
where $\pss_i$ is the stationary distribution of the original transition rate $R$
\eqa{
\sum_j R_{ij}\pss_j=0
}{st-cond}
for all $i$.
The dual transition rate is indeed a transition rate, which means that $\tlr$ satisfies both the nonnegativity condition and the normalization condition.
The former is satisfied by construction, and the latter is confirmed by using the stationary condition \eref{st-cond} as
\eq{
\sum_i \tlr_{ij}=\frac{\sum_i R_{ji}\pss_i}{\pss_j} =0.
}
In addition, the dual transition rate also has the same distribution $\pss$ as its stationary distribution:
\eq{
\sum_j \tlr_{ij}\pss_j={\sum_j R_{ji}\pss_i}=0.
}
Now we are ready to define the Hatano-Sasa excess entropy production:

\bdf{[Hatano-Sasa excess entropy production~\cite{HS01}]
Using the dual transition rate, the Hatano-Sasa entropy production rate is defined as
\eqa{
\dsgm^{\rm HS}=\sum_{i,j} R_{ij} p_j\ln \frac{R_{ij} p_j}{\tlr_{ji} p_i}.
}{dsgmHS}
}

If the system is attached to a single heat bath, then the stationary distribution is equal to the canonical distribution $\pss_i=e^{-\beta E_i}/Z$, and the HS entropy production reduces to the original entropy production.
Importantly, the Hatano-Sasa entropy production vanishes in the quasistatic limit.
Therefore, the inequality
\eq{
\sigma^{\rm HS}\geq 0
}
with $\sigma^{\rm HS}:=\int_0^\tau dt \dsgm^{\rm HS}$ is sometimes regarded as the counterpart of the second law of thermodynamics in the general stochastic processes.

The Hatano-Sasa entropy production itself is well-defined and nonnegative for any Markov jump processes regardless of the local detailed-balance condition.
On the other hand, the corresponding housekeeping entropy production $\sigma^{\rm HS, hk}$, which is the difference between the entropy production and the HS entropy production, can be negative in the case without the local detailed-balance condition.
Under the local detailed-balance condition, the housekeeping entropy production can be expressed as
\eq{
\sigma^{\rm HS, hk}=\sum_{i,j} R_{ij} p_j\ln \frac{\tlr_{ji}}{R_{ji}},
}
which can be shown to be nonnegative.
Further properties of the Hatano-Sasa entropy production are discussed in Ref.~\cite{Shibook}.

\bigskip

We next investigate the Maes-Neto\v{c}n\'{y}-type excess entropy production (MN entropy production).
Although the original paper of Maes and Neto\v{c}n\'{y}~\cite{MN14} considered this excess entropy production for overdamped Langevin systems, we here employ an analogical quantity to it defined for Markov jump processes~\cite{Kol22}.
We consider a Markov jump process with the following property:
For any edge $i$, there exists a corresponding edge $\overline{ij}$ such that the entropy production can be expressed as
\eqa{
\dsgm=\sum_{i,j} j_{ij}\ln \frac{j_{ij}}{j_{\overline{ij}} }
}{MN-condition}
with $j_{ij}=R_{ij}p_j$.
The above condition includes the local detailed-balance condition \eqref{LDB} as the case with $\overline{ij}=ji$.

We shall introduce another decomposition of the entropy production rate into the excess part $\dsgmex$ and the housekeeping part $\dsgmhk$.
In the following, we drop the time dependence of $R$ and $\bsp$ for brevity.
To explain the decomposition, we introduce the generalized Kullback-Leibler divergence defined as
\eq{
\calD(\bsv||\bsu)=\sum_i v_i \ln \frac{v_i}{u_i}+v_i-u_i
}
for any vectors $\bsv, \bsu$ with $u_i,v_i\geq 0$ including the case of $\sum_i v_i\neq \sum_i u_i$.
Note that this quantity is nonnegative owing to a simple relation $a\ln \frac ab+b-a\geq 0$.


\bigskip

\bdf{[MN-type excess entropy production rate~\cite{Kol22}]
Given the transition rate $R$ and the probability distribution $\bsp$.
Then, the MN-type excess entropy production rate is defined as
\eqa{
\dsgmex :=\inf_{R'; R'\bsp=R\bsp}\calD(\bsj'||\bar{\bsj})
}{def-excess-var}
with $j'_{ij}=R'_{ij}p_j$, where $\bar{j}_{ij}:=j_{\overline{ij}}$ and $R'$ runs all possible transition rate satisfying
\eqa{
R'\bsp=R\bsp.
}{excess-cons}
}

The constraint \eqref{excess-cons} can also be written as
\eqa{
\sum_j (j_{ij}-j_{ji})=\sum_j (j'_{ij}-j'_{ji})
}{excess-cons-j}
for all $i$.
The constraint \eqref{excess-cons} means that the time evolutions in the probability distribution $\bsp$ produced by $R$ and $R'$ are the same.

The housekeeping heat is defined by another optimization problem:

\bdf{[MN-type housekeeping entropy production rate~\cite{Kol22}]
Given the transition rate $R$ and the probability distribution $\bsp$.
Then, the housekeeping entropy production rate is defined as
\eqa{
\dsgmhk :=\inf_{\phi}\calD(\bsj||\bsj'')
}{def-hk-var}
with $j''_{ij}=R''_{ij}p_j$ and
\eqa{
R''_{ij}:=R_{ji}e^{\phi_i-\phi_j}.
}{hk-R}
}


To demonstrate the well-definedness and meaningfulness of this decomposition, Kolchinsky, {\it et al.}~\cite{Kol22} showed that the sum of these two indeed reproduces the entropy production rate
\eqa{
\dsgm=\dsgmex+\dsgmhk.
}{dsgm-decomp-var}
In addition, denoting  by $R^*$, $\bsj^*$, $R^{**}$, and $\bsj^{**}$ the optimal transition rate minimizing the right-hand sides of \eref{def-excess-var} and \eref{def-hk-var} and corresponding fluxes, respectively, these two optimizers coincide
\balign{
R^*&=&R^{**}, \nt \\
\bsj^*&=&\bsj^{**}.
} 
The latter implies that the entropy production rate $\dsgm$ is decomposed as
\eq{
\calD(\bsj||\bar{\bsj})=\calD(\bsj||\bsj^*)+\calD(\bsj^*||\bar{\bsj}).
}

\bigskip

It is shown in Ref.~\cite{Kol22} that the HS entropy production is always less than or equal to the MN entropy production
\eq{
\dsgmex\geq \dsgm^{\rm HS}.
}
In addition, the L$^1$-Wasserstein distance is known to be always larger than or equal to the (half of) total variation distance:
\eq{
\frac12 \|p-p'\|_1 \leq \calW_1(p,p').
}
In fact, the Wasserstein distance meets half of the total variation distance for a complete graph, and otherwise the former is larger than the latter in general.
Hence, the combination of the HS entropy production and the L$^1$-Wasserstein distance will lead to the strongest speed limit inequality.
However, at present, we only have speed limit inequalities with (i) the MN entropy production and the L$^1$-Wasserstein distance, or (ii) the HS entropy production and the total variation distance, and the ideal strongest inequality is elusive.
Below, we first present two types of speed limit inequalities with some extensions, and then briefly comment on why the ideal strongest speed limit inequality has not yet been addressed.

\subsection{Speed limit with MN entropy production}

We first briefly see a speed limit inequality with MN entropy production and the Wasserstein distance.
Define a current and activity with a state variable $a_i$ as
\balign{
J_a&:=&\sum_{i,j}(a_i-a_j)j_{ij}, \\
A_a&:=&\sum_{i,j}\abs{a_i-a_j}j_{ij}.
}
Notice that the above definition of $A_a$ is slightly different from the definition of $A_w$ in \eref{def-Aw}.
We denote the time integral of current by
\balign{
\calJ_a&:=& \int_0^\tau J_adt. 
}

Kolchinsky, {\it et al.}~\cite{Kol22} showed that the MN entropy production is bounded from below by $J_a$ and $A_a$ if the observable satisfies $\abs{a_i-a_j}\leq 1$ for all neighboring states $i$ and $j$:
\eqa{
\dsgmex \geq 2J_a \, {\rm arctanh} \frac{J_a}{A_a}\geq \frac{2J_a^2}{A_a}.
}{Kol-main}
Here, we used $x \, {\rm arctanh} x \geq x^2$ in the second inequality.
For completeness, we present the derivation of \eref{Kol-main} in the Appendix.A.
The Schwarz inequality leads to a trade-off inequality for arbitrary time-integrated quantities with $\abs{a_i-a_j}\leq 1$:
\eqa{
\sigma^{\rm MN}\cdot \tau\la A_a\ra_\tau=\( \int \dsgmex dt\) \( \int A_a dt\) \geq \( \int \sqrt{2}\abs{J_a} dt\) ^2 \geq 2\calJ_a^2 .
}{MN-mid1}
To cover general state quantities, not restricted to quantities with $\abs{a_i-a_j}\leq 1$, we normalize a general state variable $a$ as $a_i':=a_i/\|\Di a\|$ with $\|\Di a\|:=\max_{i,j}\abs{a_i-a_j}$ and consider the above inequality \eref{MN-mid1} with $a'$.
The obtained inequality is as follows:

\bthm{[Kolchinsky, {\it et al.}~\cite{Kol22}]
Consider a system with the condition \eqref{MN-condition}.
For a general state variable $a$, the MN entropy production is bounded from below as
\eqa{
\tau \geq \frac{2\calJ_a^2}{ \|\Di a\| \la A_a\ra_\tau\sigma^{\rm MN}}.
}{SL-Kol}
}

Let us compare the obtained inequality with \eref{main-w} by choosing the parameter at $w_{ij}=a_i-a_j$, as discussed in Sec.~\ref{s:SL-tradeoff}.
Although we do not have a simple relation between $\sigma^{\rm MN}$ and $\Pi$, we expect $\sigma\simeq \Pi$ not extremely far from equilibrium, leading to $\Pi\simeq \sigma\geq \sigma^{\rm MN}$.
On the other hand, $\la A_w\ra_\tau$ with the above choice of $w$ is less than or equal to $\|\Di a\| \la A_a\ra_\tau$.
Thus, even assuming $\sigma\simeq \Pi$, \eref{SL-Kol} and \eref{main-w} do not have a simple relation in which one is stronger than the other.

\bigskip

Setting $a$ as the optimizer $\psi$ in the Kantorovich-Rubinstein duality \eref{KR-duality}, we have a speed limit inequality with the MN entropy production.
In this reduction, we use $\|\Di \psi\|=1$, $A_\psi\leq A$, and $\int_0^\tau \abs{\calJ_\psi}dt\leq \calW_1(p(0), p(\tau))$.

\bthm{[Kolchinsky, {\it et al.}~\cite{Kol22}]
Consider a system with the condition \eqref{MN-condition}.
The MN entropy production bounds the speed of the process as
\eq{
\tau \geq \frac{2\calW_1^2(p,p')}{\overline{A}^\tau\sigma^{\rm MN}}.
}
}

\subsection{Speed limit with HS entropy production}

We next investigate speed limit inequalities with the HS entropy production.
To this end, we introduce a generalized entropy production rate and pseudo entropy production rate for a general transition rate $R'$ as
\balign{
\dsgm^{R'}&:=\sum_{i,j} R_{ij} p_j\ln \frac{R_{ij} p_j}{R'_{ji} p_i}, \\
\dPi^{R'}&:=\frac12 \sum_{i,j}\frac{(R_{ij} p_j-R'_{ji} p_i)^2}{R_{ij} p_j+R'_{ji} p_i}.
}
The original entropy production rate and pseudo entropy production rate are $\dsgm^R$ and $\dPi^R$, respectively.
The Hatano-Sasa entropy production rate is written as $\dsgm^{\tlr}$ in the above expression.
The generalized entropy production $\sigma^{R'}$ and generalized pseudo entropy production $\Pi^{R'}$ are their time integrations.

Unlike the case of the original entropy production rate, we no longer have a relation $\dsgm\geq \dPi$, and instead we have a weaker one:
\eqa{
\dsgm^{R'}\geq c^* \dPi^{R'}
}{sgm-Pi-c*}
with a constant $c^*=0.896\cdots$.
The constant $c^*$ comes from an elementary inequality
\eq{
(a-b)\ln \frac ab +b-a \geq c^*\frac{(a-b)^2}{a+b},
}
which was pointed out in Ref.~\cite{SST16}.
Remark that the constant $c^*$ is not artificial in that this type of inequality with $c^*$ achieves its equality~\cite{BHS17}.

\bigskip


We shall prove two speed limit inequalities for generalized entropy production and pseudo entropy production.

\bthm{\lb{t:st-Pi}
Consider a Markov jump process whose initial and final states are $p(0)=p$ and $p(\tau)=p'$.
Then, for any transition rate $R'$ satisfying $R'_{ii}=R_{ii}$ for all $i$, the operation time $\tau$ is bounded as
\eqa{
\tau \geq \frac{\|p-p'\|_1^2}{2\overline{A}^\tau \Pi^{R'}}.
}{main-st-Pi}
}

\bthm{\lb{t:st-sgm}
Consider a Markov jump process whose initial and final states are $p(0)=p$ and $p(\tau)=p'$.
Then, for any transition rate $R'$ satisfying $R'_{ii}=R_{ii}$ for all $i$, the operation time $\tau$ is bounded as
\eqa{
\tau \geq \frac{\|p-p'\|_1^2}{2\overline{A}^\tau \sigma^{R'}}.
}{main-st-sgm}
}

A particularly important case is $R'=\tlr$, which is treated in Ref.~\cite{VVH20}.
We can set the dual transition rate as $R'$ since
\eqa{
\tlr_{ii}=-\sum_{j(\neq i)}\frac{R_{ij}\pss_j}{\pss_i}=R_{ii}.
}{HS-diagonal}
In this case, \eref{main-st-sgm} reads
\eqa{
\tau \geq \frac{\|p-p'\|_1^2}{2\overline{A}^\tau \sigma^{\rm HS}}.
}{SL-HS}
However, the minimum of $\Pi^{R'}$ and $\sigma^{R'}$ over all possible $R'$ satisfying the constraint is achieved not at $R'=\tlr$ in general.
This means that \eref{main-st-sgm} with $R'=\tlr$ is still not optimal, and we have some room to improve inequalities.

Since we only have \eref{sgm-Pi-c*} and does not have $\dsgm^{R'}\geq \dPi^{R'}$, \eref{main-st-Pi} does not imply \eref{main-st-sgm}, and we need to prove these two inequalities separately.

\bpf{[Proof of Theorem.~\ref{t:st-Pi}]
Most of the proof follows the proof technique for Theorem~\ref{t:main-w}.

We introduce modified current $j'_{ji}(t):=R'_{ji}(t)p_i(t)$ and define three modified quantities as
\balign{
\Pi^{R'}_{ij}&=&\int_0^\tau dt \frac{(j_{ij}(t)-j'_{ji}(t))^2}{j_{ij}(t)+j'_{ji}(t)}, \\
\overline{\alpha^{R,R'}_{ij}}^\tau&=& \frac1\tau\int_0^\tau dt (j_{ij}(t)+j'_{ji}(t)), \\
\pi^{R'}_{ij}&=&\int_0^\tau dt \abs{j_{ij}(t)-j'_{ji}(t)}.
}
The Schwarz inequality leads to
\balign{
\tau \overline{\alpha^{R,R'}_{ij}}^\tau \Pi^{R'}_{ij}&=&\( \int_0^\tau dt (j_{ij}(t)+j'_{ji}(t)) \) \( \int_0^\tau dt \frac{(j_{ij}(t)-j'_{ji}(t))^2}{j_{ij}(t)+j'_{ji}(t)}\) \nt \\
&\geq&\( \int_0^\tau dt \abs{j_{ij}(t)-j'_{ji}(t)} \) ^2 \nt \\
&=&(\pi^{R'}_{ij})^2.
}

Now we use the condition on $R'$; the coincidence of the diagonal terms with $R$.
This condition leads to an important relation
\eq{
\sum_{i(\neq j)} j'_{ij}=\sum_{i(\neq j)} R'_{ij}p_j=\sum_{i(\neq j)}R_{ij}p_j=\sum_{i(\neq j)}j_{ij},
}
which implies
\balign{
A&=&\frac12 \sum_{i,j}(j_{ij}+j_{ji})=\frac12 \sum_{i,j}(j_{ij}+j'_{ji})=\frac12 \sum_{i,j}\alpha^{R,R'}_{ij}.
}
This condition also leads to the connection between the L$^1$-norm and $\pi^{R'}$:
\balign{
\|p-p'\|_1&= &\sum_i \abs{\int_0^\tau dt \frac{d}{dt}p_i } \nt \\
&\leq & \int_0^\tau dt \sum_i \abs{\frac{d}{dt}p_i } \nt \\
&=& \int_0^\tau dt \sum_i \abs{\sum_{j(\neq i)} (j_{ij}-j_{ji})} \nt \\
&=& \int_0^\tau dt \sum_i \abs{\sum_{j(\neq i)} (j_{ij}-j'_{ji})} \nt \\
&\leq &  \int_0^\tau dt \sum_i \sum_{j(\neq i)} \abs{j_{ij}-j'_{ji}} \nt \\
&=&\sum_{i\neq j} \pi^{R'}_{ij}.
}
Combining them, we arrive at the desired inequality:
\balign{
2\tau \overline{A}^\tau \Pi^{R'} =\( \sum_{i\neq j} \overline{\alpha^{R,R'}_{ij}}^\tau \) \( \sum_{i\neq j}\Pi^{R'}_{ij}\)
\geq\( \sum_{i\neq j} \pi^{R'}_{ij}\) ^2=\|p-p'\|_1^2.
}
}

\bpf{[Proof of Theorem.\ref{t:st-sgm}]
The main proof technique shown below follows Ref.~\cite{VVH20}.
Using the following elementary inequality
\eq{
a\ln \frac ab +b-a =\int_0^1 ds \frac{s(a-b)^2}{(1-s)a+sb},
}
the generalized entropy production rate can be expressed as
\eqa{
\dsgm^{R'}=\int_0^1 ds \sum_{i \neq j} \frac{s(R_{ji}p_i-R'_{ij}p_j)^2}{(1-s)R_{ji}p_i+sR'_{ij}p_j}.
}{HS-sigma-useful}
Notably, the integrand of \eref{HS-sigma-useful} is bounded from below as
\eqa{
\sum_{i \neq j} \frac{s(R_{ji}p_i-R'_{ij}p_j)^2}{(1-s)R_{ji}p_i+sR'_{ij}p_j} \geq \frac{s \( \sum_{i\neq j} \abs{R'_{ij}p_j-R_{ji}p_i} \) ^2}{\sum_{i \neq j} (1-s)R_{ji}p_i+sR'_{ij}p_j},
}{csl-gen-mid1}
which is confirmed by multiplying the denominator of the right-hand side and applying the Schwarz inequality.
We note that using $\sum_{i(\neq j)}R'_{ij}=\sum_{i(\neq j)}R_{ij}$, the denominator is shown to be equal to the activity $A$ independent of $s$:
\eq{
 \sum_{i \neq j} (1-s)R_{ji}p_i+sR'_{ij}p_j =(1-s)A+sA=A.
}
The integration of \eref{csl-gen-mid1} with $s$ from $s=0$ to $s=1$ reads a key inequality
\eqa{
\dsgm^{R'} \geq \frac1A \( \int_0^1 s ds\) \( \sum_{i\neq j} \abs{R'_{ij}p_j-R_{ji}p_i} \) ^2\geq \frac{1}{2A} \( \sum_i \Bigl|\frac{d p_i}{dt} \Bigr|\) ^2
}{second-key}
The second inequality follows from
\eq{
\sum_{j(\neq i)} \abs{R'_{ij}p_j-R_{ji}p_i}\geq \abs{ \sum_{j(\neq i)}R'_{ij}p_j-R_{ji}p_i}= \abs{ \sum_{j(\neq i)}R_{ij}p_j-R_{ji}p_i}=\abs{\frac{dp_i}{dt}}.
}
Following the same procedure as in the case of Theorem~\ref{t:st-Pi}, \eref{second-key} leads to the desired inequality.
}

\subsection{Why Wasserstein distance and Hatano-Sasa entropy production do not meet?}

As commented before, the strongest combination for a tight speed limit inequality is the Wasserstein distance and the Hatano-Sasa entropy production.
We here briefly argue why a speed limit with this combination is not easy to obtain.

As demonstrated in the previous subsection, the derivation of the speed limit inequality with the Hatano-Sasa entropy production \eqref{SL-HS} only relies on \eqref{HS-diagonal}, the coincidence of the diagonal elements of the original and dual transition rates.
Hence, if a speed limit inequality with the Wasserstein distance and the Hatano-Sasa entropy production is derived by using a similar technique, the inequality with replacing the Hatano-Sasa entropy production $\sigma^{\rm HS}$ by the generalized entropy production $\sigma^{R'}$
\eqa{
\tau \geq \frac{2\calW_1^2(p,p')}{\overline{A}^\tau\sigma^{R'}} \hspace{20pt} {\rm (Incorrect!)}
}{incorrect-SL-HS}
should also be derived with $R'$ satisfying $R'_{ii}=R_{ii}$ for all $i$.
However, this inequality has a simple counterexample.

Consider a system with states $1,2,\ldots , N$, where the transition rate is nonzero only in the form of $R_{n\pm1,n}$.
Suppose that the initial state is equipartition ($p_n=1/N$ for all $n$), and after a small time interval $\Di t$, the probability $p_1$ decreases to $1/N-\Di p$, $p_N$ increases to $1/N+\Di p$, and others are kept unchanged.
Then, the Wasserstein distance reads 
\eq{
\calW_1(p(0),p(\Di t))=\Di p(N-1).
}
To realize this transport, $(j_{n+1,n}-j_{n,n+1})\Di t=\Di p$ should be satisfied, which is realized, e.g., by setting $R_{n+1,n}=2N\frac{\Di p}{\Di t}$ and $R_{n-1,n}=N\frac{\Di p}{\Di t}$ for all $n$.
In this setting, the averaged activity reads 
\eq{
\overline{A}^\tau=3\frac{\Di p}{\Di t}\cdot \Di t =3\Di p.
}
However, by setting $R'_{n-1,n}=2N\frac{\Di p}{\Di t}$ and $R'_{n+1,n}=N\frac{\Di p}{\Di t}$ for $2\leq n\leq N-1$, and $R'_{2,1}=2N\frac{\Di p}{\Di t}$ and $R'_{N-1,N}=N\frac{\Di p}{\Di t}$, the generalized entropy production rate reads 
\eq{
\sigma^{R'}=\frac{\Di p}{\Di t}\ln 2.
}
Hence, \eref{incorrect-SL-HS} is violated for $N\geq 1+3\ln 2=3.079\cdots$.

This trouble essentially stems from the difference between the ideas behind two distances.
The Wasserstein distance cares the transport length between two microscopic states, and thus $\calW_1(p(0),p(\Di t))$ in the above case is proportional to the distance of two endpoint states, $N-1$.
On the other hand, the total variation distance cares only about the change in the probability distribution of each state, and $\| p(0)- p(\Di t)\|_1=2\Di p$ does not depend on $N$.
The condition $R'_{ii}=R_{ii}$ for all $i$ concerns the change in the probability distribution of each state, and does not care about the distance between two microscopic states, implying that this condition is compatible with the total variation distance, not the Wasserstein distance.

On the basis of this observation, if we can obtain a speed limit inequality with the Hatano-Sasa entropy production and the Wasserstein distance, we need further detailed properties of the Hatano-Sasa entropy production than the condition $\tlr_{ii}=R_{ii}$ for all $i$.

\section{Comparing speed limit inequalities for Markov jump processes and overdamped Langevin systems}\lb{s:Langevin}

\subsection{Benamou-Brenier formula and speed limit with L$^2$ wasserstein distance}

Unlike previous sections, in this section, we treat transport problems and stochastic dynamics on a continuous Euclidean space $\bbR$.
In a continuous space, the L$^2$-Wasserstein distance with the Euclid norm has a concise expression called Benamou-Brenier formula~\cite{BB00}:

\bpro{[Benamou-Brenier formula~\cite{BB00}]
For two probability distribution $p$ and $p'$ on $\bbR$, the L$^2$-Wasserstein distance between $p$ and $p'$ is written as
\eqa{
\calW_2^2(p,p')=\inf_{q(x,t), v(x,t)} \[ \int_0^1dt \int dx \abs{v(x,t)}^2 q(x,t) \],
}{BB-formula}
where the probability distribution $q(x,t)$ satisfies the boundary condition
\eq{
p(x)=q(x,0), \hspace{10pt} p'(x)=q(x,1)
}
and the time-evolution equation with the velocity field $v(x,t)$;
\eq{
\frac{\del}{\del t}q(x,t)+\frac{\del}{\del x}\cdot (v(x,t)q(x,t))=0.
}
}

This formula is suggestive of the stochastic thermodynamics of overdamped Langevin systems with uniform friction coefficient~\cite{JKO98}.
Consider a one-dimensional overdamped Langevin system given by
\eqa{
\gamma(x) \odot \dot{x}=F(x)+\sqrt{2\gamma(x)T} \odot \xi (t)
}{overL}
with a Gaussian white noise $\xi(t)$, external force $F(x)$, friction coefficient $\gamma(x)$, and temperature $T$.
The product $\odot$ is the anti-It\^{o} product~\cite{Shi22}.
Recall that the Fokker-Planck equation corresponding to this overdamped Langevin equation reads
\eqa{
\frac{\del}{\del t}p(x,t)=-\frac{\del}{\del x} (\nu (x,t)p(x,t))=-\frac{\del}{\del x} J(x,t)
}{FP}
with the local mean velocity $\nu(x,t)$ defined as
\eqa{
\nu (x,t):=\frac{1}{\gamma(x)}F(x,t)  -\frac{T}{\gamma(x)} \frac{\frac{\del}{\del x} p(x,t)}{p(x,t)}
}{def-nu}
and the local probability current 
\eq{
J(x,t):=\nu (x,t)p(x,t).
}
In addition, the entropy production rate is given by 
\eqa{
\dsgm=\int dx \frac{\gamma(x)\abs{\nu(x,t)}^2 p(x,t)}{T}=\int dx \frac{\gamma(x)\abs{J(x,t)}^2 }{Tp(x,t)}
}{Langevin-sgm}

If the friction coefficient $\gamma(x)$ is uniform (i.e., independent of $x$), the integrand of the right-hand side of \eref{Langevin-sgm} is the same as the Benamou-Brenier formula \eqref{BB-formula} besides a numerical factor.
This argument naturally leads to the speed limit inequality for overdamped Langevin systems:

\bthm{[Aurell, {\it et al.}~\cite{Aur12}]
Consider an overdamped Langevin system with a uniform friction coefficient.
We drive the state from $p=p(0)$ to $p'=p(\tau)$ in time interval $\tau$.
Then, the entropy production bounds the time interval $\tau$ from below as
\eqa{
\tau \geq \frac{\gamma\calW_2^2(p,p')}{ T\sigma}.
}{bound-Aurell}
}

In the overdamped Langevin systems, the pseudo entropy production $\Pi$ converges to the entropy production $\sigma$~\cite{Shi21}.
Keeping this point in mind, we can see a strong similarity between this speed limit \eqref{bound-Aurell} and the speed limit inequalities \eqref{ineq-Dechant} and \eqref{main-w}, as identifying the inverse of the friction coefficient $\gamma$ to the averaged activity $\overline{A}^\tau$.

On the other hand, these two speed limits have a crucial difference that the Wasserstein distance in \eref{bound-Aurell} is the L$^2$-Wasserstein distance while that in \eqref{ineq-Dechant} and \eqref{main-w} is the L$^1$-Wasserstein distance.
Thus, whether these two are essentially the same inequality with slightly different appearances or not is a controversial issue.
Dechant~\cite{Dec22} argued that these two bounds have essentially the same origin despite their several apparent differences.
Contrarily to this anticipation, below we demonstrate that these two are indeed different bounds despite their similarity.

\subsection{Speed limit with L$^1$-Wasserstein distance for overdamped Langevin systems}

To compare these two inequalities, we take the Langevin limit for Markov jump processes on discrete states (see also Ref.~\cite{Shibook} for the case with a uniform friction coefficient).
We set the transition rate of $x$ on a one-dimensional discrete lattice with lattice length $\ep$ as
\eqa{
P_{x\to x\pm \ep}:=\frac{1}{\beta\gamma (x\pm \ep/2) \ep^2}e^{\pm \beta F(x\pm \ep/2)\ep /2},
}{disc-overL}
where the argument of $F$ and $\gamma$ is determined as the middle point of the initial and the final positions; $[x+(x\pm \ep)]/2$.
We perform the Taylor expansion of transition rates in $\ep$:
\balign{
P_{x\to x\pm \ep}&=B &\( 1\pm \frac{\beta F(x)\ep}{2}+\frac{\beta^2F^2(x)\ep^2}{8}+\frac{\beta F'(x)\ep^2}{4} \right. \nt \\
&&\left.\mp\gamma \( \frac{1}{\gamma}\) '\frac{\ep}{2} - \gamma \( \frac{1}{\gamma}\) '  \frac{\beta F(x)\ep^2}{4} +\gamma \( \frac{1}{\gamma}\) '' \frac{\ep^2}{8} +o(\ep^2) \) , \\
P_{x\pm \ep\to x}&=B& \( 1\mp \frac{\beta F(x)\ep}{2}+\frac{\beta^2F^2(x)\ep^2}{8}-\frac{\beta F'(x)\ep^2}{4}  \right. \nt \\
&&\left.\mp\gamma \( \frac{1}{\gamma}\) '\frac{\ep}{2} + \gamma \( \frac{1}{\gamma}\) '  \frac{\beta F(x)\ep^2}{4} +\gamma \( \frac{1}{\gamma}\) '' \frac{\ep^2}{8} +o(\ep^2) \) .
}
with $B=1/\beta\gamma(x) \ep^2$ and $F'(x)=dF/dx$.
Plugging these relations into the master equation, we find
\balign{
\frac{d}{dt}P(x)
&=& -p(x)(P_{x\to x+\ep}+P_{x\to x-\ep})+p(x+\ep)P_{x+\ep\to x}+p(x-\ep)P_{x-\ep\to x} \nt \\
&=&B\( 1+\frac{\beta^2F^2(x) \ep^2}{8}+\gamma \( \frac{1}{\gamma}\) '' \frac{\ep^2}{8} \) (p(x+\ep)+p(x-\ep)-2p(x)) \nt \\
&&+B\( \frac{\beta F(x) \ep}{2}+\gamma \( \frac{1}{\gamma}\) '\frac{\ep}{2} \) (p(x+\ep)-p(x-\ep))\nt \\
&&-B\beta F'(x)\ep^2p(x) +B\gamma \( \frac{1}{\gamma}\) ' \beta F(x)\ep^2 p(x) +O(\ep).
}
Taking $\ep\to 0$ limit, we recover the Fokker-Planck equation \eqref{FP}:
\eq{
\frac{d}{dt}p(x)=-\frac{\del}{\del x}\frac{F(x)p(x)}{\gamma(x)}  + T \frac{\del}{\del x}\frac{1}{\gamma(x)}\frac{\del}{\del x} p(x).
}

Now we consider \eref{ineq-Dechant} and \eref{main-w} on the discrete lattice with lattice length $\ep$ and take the above continuous limit to recover the overdamped Langevin dynamics.
We assumed that the weight $w_{x\pm \ep, x}$ converges to a smooth function $w(x)$ in this limit.
Both the entropy production $\sigma$ and pseudo entropy production $\Pi$ converges to the entropy production \eref{Langevin-sgm} with coefficient $\ep$ in this limit~\cite{Shi21}.
The generalized activity $A_w$ becomes
\balign{
\lim_{\ep \to 0}\ep^2 \sum_{x,\pm} p_\ep(x)P_{x\to x\pm \ep}w^2_{x\pm \ep, x}&=&\lim_{\ep \to 0} \ep^2 \sum_{x,\pm} p_\ep(x)\frac{w^2(x)}{\beta \gamma(x)\ep^2} \nt \\
&=&2\int dx \frac{p(x)w^2(x)}{\beta \gamma(x)}.
}
In addition, the L$^1$-Wasserstein distance becomes
\eq{
\lim_{\ep \to 0}\calW_{1,w}(p,p')=\calW_{1,w(x)}(p,p').
}
Here, no normalization factor is multiplied because the distance scales as $O(1/\ep)$ while $p$ scales as $O(\ep)$.
Combining them, we arrive at the overdamped Langevin counterpart of \eref{main-w}:
\eqa{
\tau \geq \frac{\calW_{1,w(x)}^2(p,p')}{\sigma\overline{\la \frac{w^2(x)}{\beta \gamma(x)}\ra_p}^\tau},
}{Langevin-w}
where we defined $\overline{ \la f(x)\ra_p}^\tau:=\frac1\tau \int_0^\tau dt \int dx f(x)p(x,t)$.
The case of $w(x)=1$ corresponds to the overdamped Langevin counterpart of \eref{ineq-Dechant}.

As clearly seen, even with $w(x)=1$ an inhomogeneous $\gamma(x)$ accompanies $p(x,t)$-dependent friction term $\overline{\la \frac{w^2(x)}{\beta \gamma(x)}\ra_p}^\tau$ in the speed limit as the case of activity in \eref{ineq-Dechant} and \eref{main-w}.
In other words, the distribution-independent coefficient $\gamma$ in \eref{bound-Aurell} is highly special to the homogeneous system.

Even with the homogeneous friction coefficient, the Wasserstein distance remains at L$^1$-Wasserstein distance and does not become L$^2$-Wasserstein distance as \eref{bound-Aurell}.
The relation $\calW_2(p,p')\geq \calW_1(p,p')$ implies a hierarchy that \eref{bound-Aurell} can reduce an inequality similar to \eref{Langevin-w} as
\eqa{
\tau\geq \frac{\gamma \calW_1^2(p,p')}{T\sigma},
}{bound-Aurell-reduce}
while the opposite is not possible.
The advantage of \eref{bound-Aurell} lies in the specialty of the homogeneous friction coefficient.

\subsection{Equality condition}

Since we derived \eref{Langevin-w} by the reduction of the speed limit for the discrete state space \eqref{main-w}, the attainability of the equality is not clear.
To clarify the equality condition, we derive \eref{Langevin-w} directly from overdamped Langevin dynamics with weighted metric $w(x)$.

We start with the following inequality coming from the Schwarz inequality:
\balign{
\la \frac{w^2(x)}{\beta \gamma(x)}\ra_p \dsgm&=& \( \int dx\frac{w^2(x)}{\beta \gamma(x)} p(x)\) \(\int dx \frac{\gamma(x)\abs{J(x,t)}^2 }{Tp(x,t)}\) \nt \\
&\geq& \( \int dx w(x) \abs{J(x,t)}\) ^2 .
}
Consider a process with $p(x,0)=p(x)$ and $p(x,\tau)=p'(x)$, whose local probability current at position $x$ and time $t$ is denoted by $J(x,t)$.
Then, its cost of transport in terms of total variation distance is given by
\eq{
\int dt \int dx w(x)\abs{J(x,t)},
}
and its minimum is the L$^1$-Wasserstein distance.
Combining them and applying the Schwarz inequality in the time direction, we arrive at the speed limit inequality with L$^1$-Wasserstein distance:
\balign{
\tau \overline{\la \frac{w^2(x)}{\beta \gamma(x)}\ra_p}^\tau \sigma &=& \( \int dt \la \frac{w^2(x)}{\beta \gamma(x)}\ra\) \( \int dt \dsgm\) \nt \\
&\geq&\( \int dt  \int dx w(x) \abs{J(x,t)}\) ^2 \nt \\
&\geq &\calW_1^2(p, p') \lb{Langevin-SL-derive}
}

\bthm{
Consider an overdamped Langevin system described by \eref{overL} and a weighted space with $w(x)$.
We then have the following speed limit inequality
\eqa{
\tau \geq \frac{\calW_1^2(p,p')}{\sigma \overline{\la \frac{w^2(x)}{\beta \gamma(x)}\ra_p}^\tau}.
}{Langevin-SL}
}

To transform \eref{Langevin-SL} as a trade-off inequality between current and dissipation, we set $w(x)=\frac{d}{dx}a(x)$, with which $\calW_1(p,p')$ reads the change in $a$; $\la a\ra_{p'}-\la a\ra_p$.

Now we shall examine the equality condition.
By definition of the Wasserstein distance, for given $p$, $p'$, and $w(x)$, there always exists an optimal local current $J^*(x,t)$ achieving the equality of the second inequality of \eqref{Langevin-SL-derive}.
Notice that $J^*(x,t)$ fully determines the trajectory of the probability distribution $p^*(x,t)$ through \eref{FP}.
To achieve the equality of the first inequality, which comes from the Schwarz inequality, we should keep the ratio of ${w^2(x)}p(x,t)/{\beta \gamma(x)}$ and $\gamma(x)\abs{J(x,t)}^2/Tp(x,t)$ constant for all $t$ and all $x$.
This can be fulfilled by setting
\balign{
\gamma^*(x)&=&C\cdot\frac{w(x)}{\beta \abs{J^*(x,t)}}p^*(x,t) \\
F^*(x,t)&=&\gamma^*(x)\frac{J^*(x,t)}{p^*(x,t)}+T\frac{\frac{\del}{\del x} p^*(x,t)}{p^*(x,t)}
}
with some constant $C$.

As perceived from the above discussion, a nonuniform friction coefficient is inevitable to achieve the equality of the speed limit \eqref{Langevin-SL}.
This fact is important in consideration of the connection between the speed limit with the L$^2$-Wasserstein distance and the L$^1$-Wasserstein distance.
Consider an overdamped Langevin system with a homogeneous friction coefficient, where \eref{bound-Aurell} applies.
From \eref{bound-Aurell}, we can obtain the reduced inequality \eqref{bound-Aurell-reduce} by using $\calW_2(p,p')\geq \calW_1(p,p')$.
However, if the L$^2$-Wasserstein distance is strictly larger than that L$^1$-Wasserstein distance ($\calW_2(p,p')>\calW_1(p,p')$) and the equality of \eref{Langevin-SL} is achieved with an inhomogeneous friction coefficient, then the reduced inequality \eqref{bound-Aurell-reduce} never achieves its equality.
Namely, \eqref{bound-Aurell-reduce} is not tight.

For the same reason, the following inequality which is naively guessed from \eref{main-w} and \eref{bound-Aurell}
\eq{
\tau \geq \frac{\calW_2^2(p,p')}{\sigma \overline{\la \frac{w^2(x)}{\beta \gamma(x)}\ra_p}^\tau} \hspace{20pt} ({\rm Incorrect!})
}
is an incorrect inequality.
This inequality is violated when $\calW_2(p,p')>\calW_1(p,p')$ holds and the equality of \eref{Langevin-SL} is achieved.

In conclusion, we summarize the relation between the type of Wasserstein distance and the controllability of the system.
If we can control both the external force and the friction coefficient (overdamped Langevin systems) or traffic (jump processes), then L$^1$-Wasserstein distance provides a tight speed limit.
On the other hand, if we can control only the external force and the friction coefficient is fixed at a uniform value (overdamped Langevin systems), then L$^2$-Wasserstein distance provides a tight speed limit.
Since fixed average traffic in a jump process is a highly unphysical requirement, we do not expect the Markov jump counterpart of \eref{bound-Aurell}.

\section{Conclusion}

We have clarified and resolved several problems in classical speed limit inequalities, from the viewpoint of the Wasserstein distance.

In \sref{SL-w}, we have derived a speed limit inequality \eqref{main-w} with the Wasserstein distance on general weighted graphs, which reproduces trade-off relation between current and dissipation.
In this inequality, the pseudo entropy production plays a pivotal role instead of the conventional entropy production.

In \sref{SL-excess}, we have generalized speed limit inequalities \eqref{main-st-Pi} and \eqref{main-st-sgm} with the Hatano-Sasa entropy production and the total variation distance.
Through comparison with the speed limit inequality with the Wasserstein distance, we have argued why the total variation distance cannot be replaced by the Wasserstein distance.

In \sref{Langevin}, we have compared the speed limit inequality for Markov jump processes \eref{main-w} with the speed limit inequality for overdamped Langevin systems \eref{bound-Aurell}.
Curiously, though these two inequalities take the same form, the former employs the L$^1$-Wasserstein distance while the latter employs the L$^2$-Wasserstein distance.
We have derived the counterpart of \eref{main-w} to overdamped Langevin systems, \eref{Langevin-SL}, and argued that \eref{main-w} and \eref{bound-Aurell} are similar but different relations, whose difference comes from the homogeneity or inhomogeneity of the friction constant.

\ack
The author thanks Sosuke Ito for the kind communication on \cite{Kol22}.
The author also thanks Tan Van Vu for careful reading and helpful comments on the draft.
This work is supported by JSPS KAKENHI Grants-in-Aid for Early-Career Scientists Grant Number JP19K14615. 


\bigskip

{\Large \bf
\begin{center}
Appendix
\end{center}
}

\setcounter{section}{0}
\renewcommand{\thesection}{\Alph{section}}

\setcounter{equation}{0}
\renewcommand{\theequation}{\Alph{section}.\arabic{equation}}

\section{Derivation of \eref{Kol-main}}

Here, we present the derivation of \eref{Kol-main} for completeness.
The following derivation is based on Ref.~\cite{Kol22}.

We start with the following expression of the MN entropy production:
\balign{
\dsgmex=\dsgm-\dsgmhk&=&\sup_{\phi}\[ \calD(\bsj||\bar{\bsj})- \calD(\bsj||\bsj'')\] \nt \\
&=&\sup_{\phi}\[\sum_{i,j} j_{ij}(\Di \phi_{ij}+1-e^{\Di \phi_{ij}}) \],
}
where we defined $\bar{j}_{ij}=j_{ji}$, $\Di \phi_{ij}:=\phi_i-\phi_j$, and $j''_{ij}=j_{ij}e^{\Di \phi_{ij}}$.
By definition of supremum, any choice of $\phi$ in the right-hand side serves as a lower bound of $\dsgmex$.
We in particular set
\eq{
\phi_i=a_i \ln \frac{A_a-J_a}{A_a+J_a},
}
with a 1-Lipshitz variable $a$.
Using an elementary inequality
\eq{
1-e^{\lmd x}\geq \abs{x} (1-e^{{\rm sgn} (x)\lmd})
}
for $\abs{x}\leq 1$, we bound MN entropy production from below as
\balign{
\dsgmex&\geq&J_a \ln \frac{A_a-J_a}{A_a+J_a} +\sum_{i,j} j_{ij}\( 1-e^{(a_i-a_j)\ln \frac{A_a-J_a}{A_a+J_a}}\) \nt \\
&\geq &J_a \ln \frac{A_a-J_a}{A_a+J_a} +\sum_{i,j} (j_{ij}+j_{ji})\abs{a_i-a_j}  \nt \\
&&- \sum_{\la i,j\ra} (a_i-a_j) \( j_{ij} \frac{A_a-J_a}{A_a+J_a} -j_{ji} \frac{A_a+J_a}{A_a-J_a} \) \nt \\
&=&J_a \ln \frac{A_a-J_a}{A_a+J_a} +A_a-\frac{A_a+J_a}{2}\frac{A_a-J_a}{A_a+J_a} -\frac{A_a-J_a}{2} \frac{A_a+J_a}{A_a-J_a} \nt \\
&=&J_a \ln \frac{A_a-J_a}{A_a+J_a} \nt \\
&=&2J_a \, {\rm arctanh} \frac{J_a}{A_a},
}
where the summation over $\la i,j\ra$ runs over all pairs of $(i,j)$ with the order satisfying $a_i\geq a_j$.
In the first equality, we used $A_a=\sum_{\la i,j\ra} (a_i-a_j) ( j_{ij} +j_{ji})$.

\end{document}